\documentclass[prd,aps,secnumarabic,superscriptaddress,floatfix,preprintnumbers]{revtex4}
\usepackage[dvips]{graphicx,color}
\usepackage{dcolumn}
\usepackage{amsmath}
\def\sc{\mbox{\rule[-5pt]{0pt}{16pt}}}

\def\sb{\mbox{\rule{0pt}{11pt}}}

\def\sw{\mbox{\rule{24pt}{0pt}}}
\def\al{\alpha}
\def\be{\beta}
\def\ga{\gamma}
\def\de{\delta}

\def\Up{\Upsilon}
\def\Dot{\!\cdot\!}

\newcommand{\mathsym}[1]{{}}
\newcommand{\unicode}[1]{{}}

\begin{document}
\vspace{12pt}
\title{Three-loop Static QCD Potential in Heavy Quarkonia}
\author{Wayne W. Repko}
\email{repko@pa.msu.edu}
\author{Marco D. Santia}
\email{santiama@msu.edu}
\affiliation{Department of Physics and Astronomy, Michigan State
University, East Lansing, MI 48824}
\author{Stanley F. Radford}\email{sradford@brockport.edu}
\affiliation{Department of Physics, The College at Brockport, State
University of New York, Brockport, NY 14420}
\date{\today}
\begin{abstract}  
\indent We investigate the effects of including the full three-loop QCD correction to the static short distance $1/r$ potential on the spectroscopy and decays in the charmonium and upsilon systems. We use a variational technique with the full three-loop corrected potential to determine a set of unperturbed trial wave functions and treat the relativistic and one-loop corrections as perturbations. The perturbed results are compared to the subset of the charmonium and upsilon spectra using a $\chi^2$ test. This approach results in more accurate descriptions of the hyperfine splittings in both the $b\bar{b}$ and $c\bar{c}$ systems. 
\end{abstract}
\maketitle
\section{INTRODUCTION}
In an earlier publication \cite{Radford}, two of us investigated the spectrum, leptonic widths, and electromagnetic decays of the charmonium and upsilon systems. We adopted a variational approach based on determining a set of trial wave functions derived from an unperturbed Hamiltonian consisting of a relativistic kinetic energy term, a long range confining potential and a short distance $1/r$ potential that included the one-loop QCD correction. These wave functions were then used to compute the contributions of the remaining relativistic and one-loop corrections to the short distance potential using perturbation theory. The resulting masses of the various states were then compared with a subset of the experimental values using a $\chi^2$ test and the procedure was repeated until a $\chi^2$ minimum was obtained.

The model in Ref.\,\cite{Radford} yielded quite reasonable results, but there were some areas of concern. These had to do with the size of the hyperfine splittings and the values of some leptonic widths. Since both of these observables are sensitive to the values of the radial wave functions at the origin, we suspected that the trial functions, even with the inclusion of the one-loop QCD correction, were not sufficiently well determined at short distances.

To address this problem, in this paper we include the one-loop, two-loop, and three-loop corrections to the short distance $1/r$ potential in the unperturbed Hamiltonian. For the two-loop corrections to the static QCD potential we use the results of Y.~Schr\"oder, Ref.\,\cite{york1}. The three-loop results are those of A.~V.~Smirnov, V.~A.~Smirnov and M.~Steinhauser, Ref.\,\cite{SSS1,SSS2}. The two-loop and three-loop derivations were performed using the $\overline{MS}$ renormalization scheme. As discussed below, the use of the $\overline{MS}$ scheme requires a slight modification to our one-loop perturbative potential. With this modification, we can then repeat the variational calculation previously mentioned.

In Sec.\,\ref{sec:2}, we discuss the modifications of the perturbative potential associated with changing it to the $\overline{MS}$ renormalization scheme and, in Sec.\,\ref{sec:3}, we present the results for the charmonium and upsilon systems. Our conclusions are contained in Sec.\,\ref{sec:4}. Several calculational details are presented in the Appendices.
  
\section{Modified Semirelativistic Model \label{sec:2}}

We use a semi-relativistic unperturbed Hamiltonian of the form
\begin{equation}\label{unpert}
H_0=2m\sqrt{\vec{p}^{\,2}+m^2}+ Ar+V(r)\,,
\end{equation}
where $V(r)$ is the complete three-loop QCD short-range static potential \cite{york1,SSS1,SSS2}. The details of its calculation are given in Appendix A and the result is
\begin{eqnarray}
V(r) &=& -\frac{4}{3} \frac{\bar{\alpha}_S}{r} \Big[1+ \Big(a_1+2\be_0\ln(\mu'r)\Big)\frac{\bar{\alpha} _S}{4\pi}  \nonumber \\ 
& & +\Big(a_2+2(2a_1\be_0+\be_1)\ln(\mu'r)+4\be_0^2(\ln^2(\mu'r)+ \frac{\pi^2}{12})\Big) \frac{\bar{\alpha }_S^2}{(4 \pi)^2} \nonumber \\
& & +\Big(a_3+2(3a_2\be_0+2a_1\be_1+\be_2+216\pi^2)\ln(\mu'r) \nonumber \\
& &+4(2a_1\be_0^2+\frac{5}{2}\be_0\be_1)(\ln^2(\mu'r)+\frac{\pi^2}{12}) +8\be_0^3 (\ln^3(\mu'r)+\frac{\pi^2}{4}\ln(\mu'r)+2\zeta(3))\Big)\frac{\bar{\alpha }_S^3}{(4\pi)^3}\Big]\,. \label{potential}
\end{eqnarray}
where $\mu'=\mu e^\ga$, $\mu$ is the renormalization scale and $\ga=0.577216$ is Euler's constant. The values of the constants $a_i$ and $\be_i$ are given in the Appendix A. There is some question about whether or not the $216\pi^2$ contribution in the $\bar{\al}_S^3$ term of Eq.\,(\ref{potential}) should be included in the static potential \cite{BPSV,BVTS}. Our results are essentially unaffected by removing this term.

The perturbative potential has the form
\begin{equation}
H'=V_L+V_S\,.
\end{equation}
Here, $V_L$ is the order $v^2/c^2$ correction to the confining potential,
\begin{equation}
V_L=-\frac{A}{2m^2r}\vec{L}\Dot\vec{S}\,.
\end{equation}
The short distance potential $V_S$ \cite{gr1,bnt,grr},
\begin{equation}
V_S=V_{HF}+V_{LS}+V_T+V_{SI}\,,
\end{equation}
must be modified in order to be consistent with the full three-loop QCD static potential $V(r)$. This amounts to changing the renormalization scheme from the Gupta-Radford version \cite{gr} used in \cite{Radford} to the $\overline{MS}$ version used in obtaining $V(r)$. The transformation between the $\al_S$ used in \cite{Radford} and $\bar{\al}_S$ is given by
\begin{equation}
\al_S=\bar{\al}_S[1+\frac{\bar{\al}_S}{4\pi}(\frac{49}{3}-\frac{10}{9}n_f)]\,,
\end{equation}
where $n_f$ is the number of light quark flavors. From this relationship we now can write down the terms in the modified short distance potential as
\begin{eqnarray}\label{mspot}
V_{HF} &=&\frac{32\pi\bar{\al}_S\vec{S}_1\Dot\vec{S}_2}{9m^2}\left\{\left[1+\frac{\bar{\al_S}} {4\pi}(\frac{23}{3}-\frac{10n_f}{9}-3\ln\,2)\right]\de(\vec{r})
-\frac{\bar{\al}_S}{24\pi^2}(33-2n_f)\nabla^2\left[\frac{\ln\,\mu r+\ga_E}{r}\right]\right. \nonumber \\ 
&& \left.+\frac{21\bar{\al}_S}{16\pi^2}\nabla^2\left[\frac{\ln\, mr+\ga_E}{r}\right]\right\}\,, \\
V_{LS} &=& \frac{2\bar{\al}_SL\cdot S}{m^2r^3}\left \{ 1+ \frac{\bar{\al}_S}{4\pi} \left ( \frac{125}{9}-\frac{10n_f}{9} +\frac{2}{3}\Big[(33-2n_f)(\ln(\mu r)+\gamma_E-1)-12(\ln mr+\gamma_E-1) \Big] \right ) \right \}\,, \\
V_{T\;}
 &=&\frac{4\bar{\al}_S(3\vec{S_1}\Dot\hat{r}\vec{S_2}\Dot\hat{r}-\vec{S_1} \Dot\vec{S_2})}{3m^2r^3}\left\{1+\frac{\bar{\al}_S}{4\pi} \left(\frac{65}{3}-\frac{10n_f}{9}+\frac{2}{3}\left[(33-2n_f)\left(\ln\mu
r+\ga_E-\frac{4}{3}\right)\right.\right.\right.\nonumber \\
&&\left.\left.\left.-18\left(\ln mr+\ga_E -\frac{4}{3}\right)\right]\right)\right\}\,, \\
V_{SI} &=&\frac{4\pi\bar{\al}_S}{3m^2}\left\{\left[1+\frac{\bar{\al}_S}{4\pi} (\frac{43}{3}-\frac{10n_f}{9}-2\ln2)\right]\de(\vec{r}) -\frac{\bar{\al}_S}{24\pi^2}(33-2n_f)\nabla^2\left[\frac{\ln\,\mu r+\ga_E}{r}\right] -\frac{7\bar{\al}_Sm}{6\pi r^2}\right\}\,. 
\end{eqnarray}

\section{Results \label{sec:3}}

In the process of treating $V_L$ and $V_S$ perturbatively, we compared the retention of the $\de$-function terms in $V_{HF}$ and $V_{SI}$ with the practice of `softening' it by using a sharply peaked, differentiable function. This is a departure from \cite{Radford}. With the modified $H_0$, we found that the softened $\de$-function was preferable in the overall $c\bar{c}$ fit while $b\bar{b}$ spectra and hyperfine splitting \cite{babar1,babar2,cleo} were better described by retaining the $\de$-function behavior. Following this procedure, the final parameters that resulted from the variational calculation are shown in Table \ref{params}.
\begin{table}[h]
\centering 
\begin{tabular}{lcc}
\toprule
\sw &\multicolumn{1}{c}{\sb $c\bar{c}$\sw}&\multicolumn{1}{c}{
\sb $b\bar{b}$} \\
\hline
\sc$A$ (GeV$^2$)\mbox{\rule{12pt}{0pt}}  & 0.132\sw  & 0.171 \\
\hline
\sc $\bar{\al}_S$ & 0.276\sw  & 0.184 \\
\hline
\sc $m_q$ (GeV)   & 1.41\sw   & 4.64  \\
\hline
\sc $\mu$ (GeV)   & 1.79\sw  & 6.10  \\
\botrule
\end{tabular}
\caption{Fitted Parameters for the $c\bar{c}$ and
$b\bar{b}$ systems\label{params}}
\end{table}

\subsection{Charmonium System}
The results for the spectrum of the charmonium system are shown in Table \ref{ccspec}. We set $n_f$ to 3 for the calculations. 
\begin{table}[h]\centering
\begin{tabular}{ldd} \toprule
\multicolumn{1}{c}{\sc}$m_{c\bar{c}}$\,(MeV)  &\multicolumn{1}{c}{Theory}  & \multicolumn{1}{c}{ Expt } \\
\hline
\sb$\eta_c(1S)^*$\mbox{\rule{12pt}{0pt}}   & 2980.4  & 2980.4\pm 1.2 \\
\hline
\sb$J/\psi (1S)^*$     & 3097.4   & 3096.916\pm 0.011  \\
\hline
\sb$\chi_{c\,0}(1P)^*$ & 3415.9   & 3414.74\pm 0.35  \\
\hline
\sb$\chi_{c\,1}(1P)^*$ & 3506.9   & 3510.77\pm 0.07 \\
\hline
\sb$\chi_{c\,2}(1P)^*$ & 3558.6   & 3556.20\pm 0.09 \\
\hline
\sb$h_c(1P)$           & 3526.6   & 3525.93\pm 0.27 \\
\hline
\sb$\eta_c(2S)$        & 3602.0   & 3638.0\pm 4.0 \\
\hline
\sb$\psi(2S)^*$        & 3685.2   & 3686.093\pm 0.034 \\
\hline
\sb$\psi(1D)$          & 3804.9   & 3771.1\pm 2.4  \\
\hline
\sb$1^3D_2$            & 3824.2  &                \\
\hline
\sb$1^3D_3$            & 3832.4   &                \\
\hline
\sb$1^1D_2$            & 3824.5   &                \\
\hline
\sb$\chi_{c\,0}(2P)$   & 3841.5   &                \\
\hline
\sb$\chi_{c\,1}(2P)$   & 3933.3   &                \\
\hline
\sb$\chi_{c\,2}(2P)$   & 3988.0  & 3929.0\pm 5.4  \\
\hline
\sb$h_c(2P)$           & 3954.6   &                \\
\hline
\sb$\eta_c(3S)$      & 4010.8    &                  \\
\hline
\sb$\psi(3S)$        & 4083.5    & 4039.0\pm 1.0    \\
\hline
\sb$\psi(2D)^*$      & 4154.4    & 4153.0\pm 3.0    \\
\hline
\sb$2^3D_2$          & 4177.7    &                  \\
\hline
\sb$2^3D_3$          & 4190.8    &                  \\
\hline
\sb$2^1D_2$          & 4179.5    &                  \\
\botrule
\end{tabular}
\caption{Results for the $c\bar{c}$ spectrum are shown. The states denoted by a $^*$ are used in the fitting procedure. In order to account higher order corrections, we used a 2.5 MeV error offset in the fit. All experimental data are taken from \protect\cite{pdg}.}\label{ccspec}
\end{table}
The description of the $c\bar{c}$ spectrum with the modified non-perturbative wave functions and a softened hyperfine interaction is consistent with our earlier calculations \cite{Radford}. The result of $117.0$ MeV for $J/\psi(1S)-\eta_C(1S)$ splitting compares favorably with the $116.6\pm 1.2$ MeV experimental value. The corresponding $\psi(2S)-\eta_C(2S)$ result of $83.2$ MeV remains significantly larger than the experimental value of $48.1\pm 4.0$ MeV. On the other hand, the three-loop unperturbed wave functions provide a better description of the leptonic widths, shown in Table\,\ref{cwid}. The dipole decay rates, which depend on the matrix elements of $r$, are not particularly affected by the changes in the short distance behavior associated with the inclusion of the three-loop corrections.

\begin{table}[h]\centering
\begin{tabular}{ldd} \toprule
\multicolumn{1}{l}{\sc}$\Gamma_{e\bar{e}}$\,(keV)  &\multicolumn{1}{c}{Theory} & \multicolumn{1}{c}{ Expt } \\

\hline
\sb$\psi(1S)$\mbox{\rule{24pt}{0pt}}     & 5.78   & 5.55\pm0.14  \\
\hline
\sb$\psi(2S)$     & 2.98   & 2.48\pm 0.06  \\
\hline
\sb$\psi(3S)$     & 2.21   & 0.86\pm 0.07 \\
\hline
\sb$\psi(1D)$     & 0.05   & 0.242\pm 0.030  \\
\botrule
\end{tabular}
\caption{Leptonic widths of the $\psi(nS)$ and the $\psi(1D)$ states are shown. The $\psi(nS)$ widths include the QCD correction factor $(1-16\bar{\al}_S/3\pi)$ and the relativistic correction described in \cite{Radford}.}\label{cwid}
\end{table}

\subsection{Upsilon System}
The results for the $b\bar{b}$ spectrum, computed using $n_f=4$, are shown in Table \ref{bottomspec}. For the levels that are not particularly sensitive to the short distance behavior of the wave functions, the results are compatible with Ref.\,\cite{Radford}. The main improvement is seen in the $\Upsilon(1S)-\eta_b(1S)$ hyperfine interval \cite{rr1,Mei,Gia,Miz}, which has increased from $47.0$ MeV to $62.0$ MeV, much closer to experimental value of  $62.3\pm 3.2$ MeV.  The $2S$ and $3S$ hyperfine intervals are also larger than those in \cite{Radford}. Our calculation is also consistent with the recent discovery of evidence for a $3^3P_J$ multiplet from the ATLAS detector \cite{ATLAS,KR,MZ,rr2,Dib}. The leptonic widths are given in Table \ref{bwid}. \begin{table}[h]\centering
\begin{tabular}{lddd} \toprule
\multicolumn{1}{l}{\sc}$\!\!m_{b\bar{b}}$\,(MeV)  &\multicolumn{1}{c}{Theory}  & \multicolumn{1}{c}{ Expt } \\
\hline
\sb$\eta_b(1S)$\mbox{\rule{12pt}{0pt}}   & 9399.6\sw  & 9398.0\pm 3.2 \\
\hline
\sb$\Up(1S)^*$         & 9461.6   & 9460.30\pm 0.26  \\
\hline
\sb$\chi_{b\,0}(1P)$   & 9872.2   & 9859.44\pm 0.52  \\
\hline
\sb$\chi_{b\,1}(1P)^*$ & 9893.2   & 9892.78\pm 0.40  \\
\hline
\sb$\chi_{b\,2}(1P)^*$ & 9906.4   & 9912.21\pm 0.40  \\
\hline
\sb$h_b(1P)$           & 9898.5   &                  \\
\hline
\sb$\eta_b(2S)$        & 9969.6   &                  \\
\hline
\sb$\Up(2S)$           & 10006.4  & 10023.26\pm 0.31 \\
\hline
\sb$\Up(1D)$           & 10149.3  &                  \\
\hline
\sb$1^3D_2$            & 10153.7  & 10161.1\pm 1.7   \\
\hline
\sb$1^3D_3$            & 10156.1  &                 \\
\hline
\sb$1^1D_2$            & 10154.0  &                 \\
\hline
\sb$\chi_{b\,0}(2P)^*$ & 10238.7  & 10232.5\pm 0.6  \\
\hline
\sb$\chi_{b\,1}(2P)^*$ & 10257.0  & 10255.46\pm 0.55\\
\hline
\sb$\chi_{b\,2}(2P)^*$ & 10268.9  & 10268.65\pm 0.55\\
\hline
\sb$h_b(2P)$           & 10261.8  &                  \\
\hline
\sb$1^3F_2$            & 10352.5  &                 \\
\hline
\sb$1^3F_3$            & 10353.1  &                 \\
\hline
\sb$1^3F_4$            & 10352.4  &                 \\
\hline
\sb$1^1F_3$            & 10352.7  &                 \\
\hline
\sb$\eta_b(3S)$        & 10328.4  &                  \\
\hline
\sb$\Up(3S)^*$         & 10351.5  & 10355.2\pm 0.5   \\
\hline
\sb$\Up(2D)$           & 10447.9  &                  \\
\hline
\sb$\chi_{b\,0}(3P)$ & 10527.2   &       \\
\hline
\sb$\chi_{b\,1}(3P)$ & 10544.3   &      \\
\hline
\sb$\chi_{b\,2}(3P)$ & 10555.6   &      \\
\hline
\sb$\langle 3^3P_J\rangle$ & 10548.7   & 10534.0\pm 9.0 \\
\botrule
\end{tabular}
\caption{Results for the $b\bar{b}$ spectrum are shown. The states denoted by a $^*$ are used in the fitting procedure. In order to account higher order corrections, we used a 2.5 MeV error offset in the fit. All experimental data are taken from \protect\cite{pdg}. The notation $\langle 3^3P_J\rangle$ denotes the spin average of the $3^3P_J$ levels.}\label{bottomspec}
\end{table}

\begin{table}[h]\centering
\begin{tabular}{ldd} \toprule
\multicolumn{1}{l}{\sc}$\Gamma_{e\bar{e}}$\,(keV)  &\multicolumn{1}{c}{Theory}  &\multicolumn{1}{c}{ Expt } \\

\hline
\sb$\Upsilon(1S)$\mbox{\rule{24pt}{0pt}}     & 1.140   & 1.340\pm0.018  \\
\hline
\sb$\Upsilon(2S)$     & 0.579   & 0.612\pm0.011  \\
\hline
\sb$\Upsilon(3S)$     & 0.439   & 0.443\pm0.008 \\
\hline
\sb$\Upsilon(4S)$     & 0.356   & 0.272\pm0.029  \\

\botrule
\end{tabular}
\caption{Leptonic widths of the $\Upsilon(nS)$ states are shown. The $\psi(nS)$ widths include the QCD correction factor $(1-16\bar{\al}_S/3\pi)$ and the relativistic correction described in \cite{Radford}.}\label{bwid}
\end{table}

\section{Conclusions}\label{sec:4}

The inclusion of the full three-loop corrections to the short distance static potential used in the determination of the unperturbed trial wave functions improves description of $c\bar{c}$ and $b\bar{b}$ observables that are sensitive to the behavior of the wave function near $r\to 0$. This enabled us to retain the $\de$-function terms in the perturbative potential when treating the upsilon states and account for the larger ground state hyperfine splittings in both the charm and upsilon systems. These improvements can be obtained without a dramatic change in the descriptions of the overall $c\bar{c}$ and $b\bar{b}$ spectra.
\begin{acknowledgements}
We would like to thank Nora Brombilla for helpful communications. This work was supported in part by the National Science Foundation under Grant PHY 1068020.
\end{acknowledgements}
\appendix
\section{Three-loop short distance potential calculational details}

The momentum space version of the complete three-loop short distance potential, $V(\vec{k}^2)$, in terms of $\overline{\al}_S$ can be found in references \cite{york1,SSS1,SSS2}. Explicitly, the expression for $V(\vec{k}^2)$ is 
\begin{equation}\label{mompot}
V(\vec{k}^2)=-\frac{4}{3}\frac{4\pi}{\vec{k}^2}\overline{\al}_S\left[1+ c_1(\vec{k}^2/\mu^2)\frac{\overline{\al}_S}{4\pi}+ c_2(\vec{k}^2/\mu^2)\frac{\overline{\al}^2_S}{(4\pi)^2}+ c_3(\vec{k}^2/\mu^2)\frac{\overline{\al}^3_S}{(4\pi)^3}\right]\,,
\end{equation}
where the $c_i(x)$ are given by \cite{Note}.
\begin{eqnarray}\label{ci's}
c_1(x) &=& a_1-\be_0\ln(x)\, \\[6pt]
c_2(x) &=& a_2-(\be_1+2\be_0\,a_1)\ln(x)+\be_0^{\,2}\ln^2(x)\, \\
c_3(x) &=& a_3-(3a_2\be_0+2a_1\be_1+\be_2+216\pi^2)\ln(x)+(3a_1\be_0^2 +\frac{5}{2}\be_0\be_1)\ln^2(x)-\be_0^3\ln^3(x)\,.
\end{eqnarray}
In order to calculate the coordinate space potential $V(r)$, 
\begin{equation}\label{V(r)}
V(r)=\frac{1}{(2\pi)^3}\int\!\!d^3k\,V(\vec{k}^2)e^{i\vec{k}\cdot\vec{r}}\,,
\end{equation}
it is necessary to evaluate infrared integrals of the form
\begin{equation}
I_n(r) = \frac{4\pi}{(2\pi)^3}\int\! d^3k\,\frac{\ln^n(\vec{k}^2/\mu^2)}{\vec{k}^2}\,e^{i\vec{k}\cdot\vec{r}}= 
\frac{2^{n+1}}{r\pi}\int_0^\infty\!\!dx\frac{\ln^n(x)}{x}\sin(\mu rx)\,.
\end{equation}
Using Mathematica, the values of the relevant $I_n(r)$ are
\begin{equation}\label{In}
I_0=\frac{1}{r}\,,\quad I_1=-\frac{2}{r}\ln(\mu'r)\,,\quad I_2= \frac{4}{r}\left(\ln^2(\mu' r)+\frac{\pi^2}{12}\right)\,, \quad I_3= -\frac{8}{r}\left(\ln^3(\mu'r)+\frac{\pi^2}{4}\ln(\mu'r)+ +2\zeta(3)\right)\,.
\end{equation}
With these results in Eq.\,(\ref{V(r)}), Eq.\,(\ref{potential}) follows.

The parameters $a_i$ and $\be_i$ in Eq.\,(\ref{potential}) are \cite{york1,SSS1,SSS2}
\begin{eqnarray}
a_1 &=& \frac{31}{3}-\frac{10}{9}n_f\,, \\
a_2 &=& \frac{4343}{18}+36\pi^2+66\zeta(3)-\frac{9}{4}\pi^4-(\frac{1229}{27} +\frac{52}{3}\zeta(3))n_f+\frac{100}{81}n_f^2\,, \\
a_3 &=& 13432.6-3289.91n_f+185.99n_f^2-1.37174n_f^3\,, \\
\be_0 &=& 11-\frac{2}{3}n_f\,, \\
\be_1 &=& 102 -\frac{38}{3}n_f \\
\be_2 &=& \frac{2857}{2}-\frac{5033}{18}n_f+\frac{325}{54}n_f^2\,.
\end{eqnarray}

\section{Calculational Details}
In the determining the trial wave functions, we use a variational approach as described in \cite{Radford}. The trial function is given by
\begin{equation} 
\psi_{j\ell s}^m(\vec{r}) = \sum_{k=0}^n C_k\left( r/R \right )^{k+\ell}e^{-r/R}{\cal Y}_{j\ell s}^m(\Omega)\,. 
\end{equation}
The calculation of the radial matrix elements of Eq.\,(\ref{potential}) using this trial function involves both Gamma functions and Polygamma functions of $z=k+k'+2\ell$ and $a=\mu'R/2$. The integrals that arise are of the form
\begin{equation}
J_n(z,a)=\frac{R^2}{2^z}\int_0^\infty\!\!dtt^{z-1}e^{-t}\ln^n(a\,t)\,,
\end{equation}
and those encountered are 
\begin{eqnarray}
J_0(z,a) &=& \frac{R^2}{2^z}\Gamma(z)\,, \\
J_1(z,a) &=& \frac{R^2}{2^z}\Gamma\left(z)(\ln(a)+\psi(z)\right)\equiv \Gamma(z)\phi(z,a)\,,\\ 
J_2(z,a) &=& \frac{R^2}{2^z}\Gamma(z)\left(\phi^2(a,z)+\psi'(z)\right)\,, \\ J_3(z,a) &=& \frac{R^2}{2^z}\Gamma(z)\left(\phi^3(a,z)+3\phi(a,z)\psi'(z)+\psi''(z)\right)\,,
\end{eqnarray}
where $\psi(z)=\Gamma'(z)/\Gamma(z)$. The matrix element $V(k,k')$ of the full three-loop short distance potential is then
\begin{eqnarray}\label{Vkk'}
V(k,k') &=& -\frac{4\bar{\al}_S}{3}\frac{R^2\Gamma(z)}{2^z}\Big[1+\big(a_1 +2\be_0\phi(a,z)\big)\frac{\bar{\al}_S}{4\pi}
+\big(a_2+2(2a_1\be_0+\be_1)\phi(a,z)\nonumber \\ &&+4\be^2_0(\phi^2(a,z)+\psi'(z)+\pi^2/12)\big)\frac{\bar{\al_S}^2}{(4\pi)^2} \nonumber +\big(a_3+2(3a_2\be_0+2a_1\be_1+\be_2+216\pi^2)\phi(a,z) \nonumber \\[4pt]
&&+4(3a_1\be^2_0+ \frac{5}{2}\be_0\be_1)(\phi^2(a,z)+\psi'(z)+\pi^2/12) \nonumber \\[4pt]
&&+8\be^3_0(\phi^3(a,z)+3\phi(a,z)\psi'(z)+\psi''(z)+\frac{\pi^2}{4}\phi(a,z)+ 2\zeta(3))\big)\frac{\bar{\al_S}^3}{(4\pi)^3}\Big]\,.
\end{eqnarray}
These matrix elements, together with the matrix elements of the remaining terms in Eq,\,(\ref{unpert}), define the variational equation
\begin{equation}
\sum_{k'=0}^n H_0(k,k')C_{k'}=E\sum_{k'=0}^n N(k,k')C_{k'}\,,
\end{equation}
that determines the unperturbed wave functions and energies. Here $N(k,k')$ is the overlap integral of the (non-orthogonal) radial wave functions and we typically use $n=12$.

\end{document}